\documentclass[review]{elsarticle}

\usepackage{hyperref}
\usepackage{chngcntr}
\usepackage{xcolor}

\journal{Journal of \LaTeX\ Templates}

\begin{document}

\begin{frontmatter}

\title{SONODYNAMIC EFFECT IN A375 MELANOMA CELLS WITH CHLORIN E6 INDUCED BY 20 kHz ULTRASOUND\tnoteref{mytitlenote}}

\author{Antonina Dadadzhanova\fnref{mymainaddress}\corref{firstcorr}}
\author{Ekaterina Kolesova  \fnref{mymainaddress}}
\author{Vladimir Maslov \fnref{mymainaddress}}
\author{Eliz Amar – Lewis   \fnref{mysecondaryaddress}}
\author{Riki Goldbart  \fnref{mysecondaryaddress}}
\author{Tamar Traitel  \fnref{mysecondaryaddress}}
\author{Joseph Kost   \fnref{mysecondaryaddress}}
\author{Anna Orlova  \fnref{mymainaddress}\corref{secondcorr}}

\cortext[firstcorr]{Corresponding author: Antonina Dadadzhanova, e-mail: fadeeva.antonina95@gmail.com}

\cortext[secondcorr]{Corresponding author: Anna Orlova, e-mail: a.o.orlova@gmail.com}

\address[mymainaddress]{ITMO University, 49 Kronverksky Ave., 197101, St. Petersburg, Russia }
\address[mysecondaryaddress]{Ben-Gurion University of the Negev, David Ben Gurion Blvd, 84105, Beer-Sheva, Israel}

\begin{abstract}
In this research, combined effect of chlorin e6 and noninvasive 20 kHz, low-powered ultrasound on melanoma cells is proposed for the first time. The proposed combination incorporates the application of safe low-frequency ultrasound that is used for transdermal drug delivery. We demonstrate that impact of the 20 kHz ultrasound on chlorin e6 leads to strong sonodynamic effect in melanoma cells. We also show that chlorin e6 keeps its monomeric form in the cells and does not aggregate after exposure to ultrasound. We propose a two-step model where sonoluminescence serves as an internal light source for the chlorin e6 photoactivation. 
\end{abstract}

\begin{keyword}
\texttt{Ultrasound \sep cancer \sep sonodynamic therapy\sep chlorin e6}\

\end{keyword}

\end{frontmatter}

\section{Introduction}

 Currently, the spread of cancer is rampant, and the search for new methods for the effective treatment of cancer is a socially significant challenge. Several traditional ways are known to treat cancer, such as chemotherapy, surgical removal, hormone therapy, radiation therapy; each of them has reported side effects \cite{jones2007epigenomics, abdel2016photodynamic, plaetzer2009photophysics}. Photodynamic therapy (PDT) is one of the most sparing and at the same time effective methods for the diagnosis and treatment of cancer \cite{henderson2020photodynamic, lan2019photosensitizers,dos2019photodynamic}. PDT is based on the generation of reactive oxygen species (ROS) by photoexcitation of photosensitizer molecules. The distinctive ability of photosensitizer is its accumulation mostly in the tumor tissues. Once exposed to visible light, a photosensitizer molecule transfers into an excited state. The collision of an excited photosensitizer with surrounding medium leads to the generation of ROS \cite{yang2019reactive, yardeni2019host}. ROS is capable of triggering the mechanisms of cancer cells apoptosis or necrosis \cite{simon2000role, fortes2012heme}. Tetrapyrrole molecules can be considered as an ideal material to develop an approach for cancer visualization and treatment. This possibility exists as tetrapyrrole  molecules as a photosensitizer selectively accumulate in cancer tissues \cite{taquet2007phthalocyanines} and efficiently generate singlet oxygen, which can be produced by transferring tetrapyrrole to its excited state. Despite the advantages of PDT over traditional cancer therapy methods, it possesses a significant drawback that limits its wide application in clinical practice. Biological tissues strongly absorb visible radiation, which is necessary for transferring the sensitizer to an excited state. This fact reduces the application of this method exclusively to the treatment of epithelial forms of cancer and small tumors \cite{agostinis2011photodynamic}. 
Thus, several approaches to overcome this hurdle are now being considered. The first is to reveal a way to create an internal source for photoexcitation of the tetrapyrrole molecule in the very tumor \cite{mao2017chemiluminescence, magalhaes2016chemiluminescence}. The second one is to find an alternative way to transfer the tetrapyrrole molecule to a triplet state, so that it can generate ROS.
Some research groups have recently demonstrated that ultrasound (US) at frequencies of 0.88 - 3 MHz and intensities of 0.1 - 2.4 W/cm$^2$ can activate traditional photosensitizes, like tetrapyrrole molecules or titania nanoparticles, in cancer cells \cite{you2016ros, xu2016sonodynamic, wang2011detection}. It has opened an avenue to a new non-invasive and extremely efficient approach called Sonodynamic Therapy (SDT) for cancer treatment, especially for inhomogeneous, large and deeply localized tumors. SDT effect is more complex as it results from the combination of different mechanisms. Currently, there are two most likely mechanisms to explain the sonodynamic effect on cancer cells, i.e. cavitation and ROS generation. Cavitation is the process of formation and collapse of bubbles producing colossal energy; the released energy is converted into chemical or mechanical forms, which leads to cell destruction \cite{yasui2018acoustic}. It has been shown that the use of nanoparticles, or their aggregates at appropriate amount and size, causes an increase in the yield of the sonochemical reaction. In this case, the addition of nanoparticles can increase the number of cavitation bubbles, thus enhancing the yield of the sonochemical reaction for US frequencies at 20 kHz and higher \cite{tuziuti2005correlation, tuziuti2004effect}.
The mechanism of ROS generation implies the interaction of US with sonosensitizers. US can produce sonoluminescence that can serve as an internal light source to excite sensitizer species \cite{canaparo2020bright, ohmura2011sonodynamic, beguin2019direct}. Recently, it has been demonstrated that traditional photosensitizers exposed to US with a frequency of 1.866 MHz and intensity of 1.5 W/cm$^2$ can lead to strong apoptotic and necrotic effects in cancer cells \cite{giuntini2018insight}.
Despite obvious benefits, like deep penetration in the human body, SDT has some side effects because the US frequencies used in SDT can destroy normal cells and tissues \cite{miller2012overview}. At the same time, the treatments by low-intensity US at frequency of 20 - 100 kHz, which have no damage to human tissues, are used for transdermal drug delivery in a wide range of diseases \cite{park2014sonophoresis, katz2004rapid, mitragotri2004low, bhatnagar2016exploitation, seto2010effects}. In addition, the penetration depth of US at a frequency of 20 kHz is much greater than that of US with a frequency of 1 MHz. \emph{S.Mitragotri}  studied the influence of 20 kHz on the integrity of the epidermis and other living tissues and showed that no cell damage has been detected \cite{mitragotri2006transdermal}. In fact, 20 kHz US allows to increase the concentration of drugs in cells and tissues without triggering the mechanisms of cells apoptosis or necrosis \cite{tezel2004topical}. 
To the best of our knowledge, 20 kHz US has never been studied in SDT applications up to now. It is surprising because it is a well-known fact that low-intensity 20 kHz US can generate sonoluminescence \cite{didenko2000multibubble, ji2018multibubble}. Therefore, we hypothesized that low-frequency  US can be potentially used as an internal energy source to excite molecular and nanostructured sensitizers in cancer cells and tumors. We, therefore, study the impact of low-intensity 20 kHz US on a photosensitizer in melanoma cells. The second-generation photosensitizer chlorin e6 (Ce6) was applied as a sensitizer. Ce6 is actively used as an effective drug for PDT of cancer due to its ability to generate singlet oxygen (SO) \cite{edge2018singlet, bashkatov2005optical, bharathiraja2017chlorin}. Unlike other tetrapyrrole molecules, Ce6 is not toxic to the body  \cite{paszko2011nanodrug}. Moreover, Ce6 is quickly removed from healthy tissues in comparison to other photosensitizers \cite{chin2009vivo}. We demonstrate here that monomers of Ce6 loaded into melanoma cells can be efficiently activated by 20 kHz US that triggers the destruction of melanoma cells.  We also show that short-time treatment of melanoma cells with Ce6 by 20 kHz US obtains the SDT effect that is comparable with the PDT effect for the same Ce6 concentration and the same incident energy dose. The difference between the efficiency of SDT and PDT may be related to the limited penetration depth of light as a result of PDT. This is the first demonstration of the efficacy of 20 kHz US in SDT.  
\section{Material and methods}

\subsection{Chemical reagents}

Dimethyl sulfoxide (DMSO) and phosphate buffer solution (PBS) were purchased from Sigma-Aldrich Inc. RPMI-1640 medium, fetal bovine serum (FBS), penicillin-streptomycin, L-Glutamine, HEPES, Nystatin and Sodium pyruvate were purchased from Biological Industries. Presto Blue assay was purchased from Fisher Scientific. Chlorin e6 (Ce6) was purchased from Frontier Scientific (USA). Bi-distilled water was used throughout the experiments. All reagents were used without further purification. 
\subsection{Preparation of Chlorin e6 solution}
Ce6 powder was dissolved in PBS solution at a concentration of 2$\cdot10^{-5}$ M.
\subsection{Cell culture}
Melanoma A375 was used as a cell line. RPMI-1640  was used as a nutrient medium, containing 10$\%$ fetal bovine serum (FBS), 1$\%$ penicillin-streptomycin, 1$\%$ L-Glutamine, 0.175$\%$ HEPES, 0.075$\%$ Nystatin, and 1$\%$ Sodium pyruvate. The cells were kept in an incubator (5$\%$  CO$_{2}$) and every two days were  washed twice with PBS solution, and a new nutrient medium was added. Ce6 (dissolved in PBS) (5$\%$ of medium volume) were added to cells in RPMI-1640 medium, so each plate's units contained (950 $\mu$L) medium and 50 $\mu$L Ce6 solutions with concentration 2$\cdot10^{-5}$ M. Ce6 concentration in the medium was kept\ $10^{-6}$\ M for all experiments.
\subsection{Cell viability assessment}
Presto Blue test assay (Fisher Scientific) was used as a test for determining the viability of cells. Cell's viability is assessed before and after cells exposure to light or US. Presto Blue is a resazurin-based chemical sensor. The addition of resazurin to the cells leads to modification of the reagent by the reducing medium of viable cells. The  reagent  turns  red  and  develops  strong  sensor  photoluminescence (PL). This color change can be detected by measuring fluorescence or absorption spectra. Presto Blue was added to the cells seeded on the plates in the amount of 10\% of the total nutrient medium, then plates were kept in the incubator for 10 minutes. The number of living cells was estimated by measurement of PL intensity of Presto Blue using Infinite M200 (Tecan) spectrophotometer.

\subsection{Spectroscopic measurements}
Absorption, PL and PL excitation spectra of Ce6 molecules in cells were measured and analyzed before and after exposure to 20 kHz US. The measurements were performed on multifunctional microplate reader Tecan Infinite M200.

\subsection{Formation of Ce6 aggregates}
To obtain solutions with the monomeric form and  aggregates of Ce6, pure PBS solution and PBS diluted in bidistilled water with the ratio of 1:500 were prepared.
Concentration of Ce6 was 2$\cdot10^{-5}$ M in each solution. Ce6 dissolved in PBS at a
concentration of 2$\cdot10^{-5}$ M was used as a reference sample.

\subsection{PDT test}
For the photodynamic test, A-375 melanoma cells were seeded in a 12-well plate (4$\cdot 10^5$ cells/well). After 24 hours, the cells were washed twice with PBS (pH= 7.4) and a Ce6 solution at a concentration of 2$\cdot10^{-5}$ M (in PBS) was added to the cells in the amount of 5$\%$ of the total volume of the nutrient medium. After 24 hours of incubation at a temperature of 37$^{\circ}$C (with 5$\%$ CO$_{2}$), the cells were washed three times with a solution of PBS and nutrient medium was added. This protocol of washing melanoma cells guarantees that all Ce6 molecules that did not penetrate into the cells were washed off with PBS from the  plate  walls  or  the  cell  surface. 
We used 650 nm LED with a power of 20 mW to study PDT on A375 cell line. The cells for PDT experiments were incubated with Ce6 for 4, 24 and 72 hours, and then subsequently irradiated by the LED for 117 seconds. An estimation of PDT effect, produced by Ce6 cells, was performed as a ratio of cell viability before and an hour after irradiation of samples. Cells incubated with Ce6 at the same conditions but without LED irradiation were used as a control for the PDT effect evaluation. Cells without Ce6 irradiated by the LED at the same conditions were used as control of light influence on cell viability. We also checked the viability of the cells grown at the same conditions with light exposure and Ce6 as a control of viability of the cell line. Cells were moved to an incubator 1 hour after light exposure, then cells were washed three times with PBS, filled with fresh medium and their viability was checked using Presto Blue test.

\subsection{SDT test}
Samples were prepared in the same way as PDT test. Ultrasonic Q-SONICA plate-horn system was used in SDT test of A375 cell line. The US exposure time was 10 seconds. The intensity of US was 0.083 W/cm$^2$ per well. Estimation of SDT effect of Ce6 on cells was performed in the same way as PDT test, i.e. as ratio of cell viability before and after the action of US on samples.

\subsection{Statistical analysis}
 All data were expressed as mean ± standard deviation. The statistical analyses were performed using GraphPad Prism 9 software, by using one-way analysis of variance (ANOVA). Asterisk (*), sharp ($^{\sharp}$) and ampersand ($^{\&}$) denote statistical significance between bars (p$<$0.05, p$<$0.01, p$<$0.001, p$<$0.0001).

\section{Results and Discussion}
\subsection{SDT test on A375 cells}

\begin{figure}[h]
	\centering
	\includegraphics[height=6 cm ]{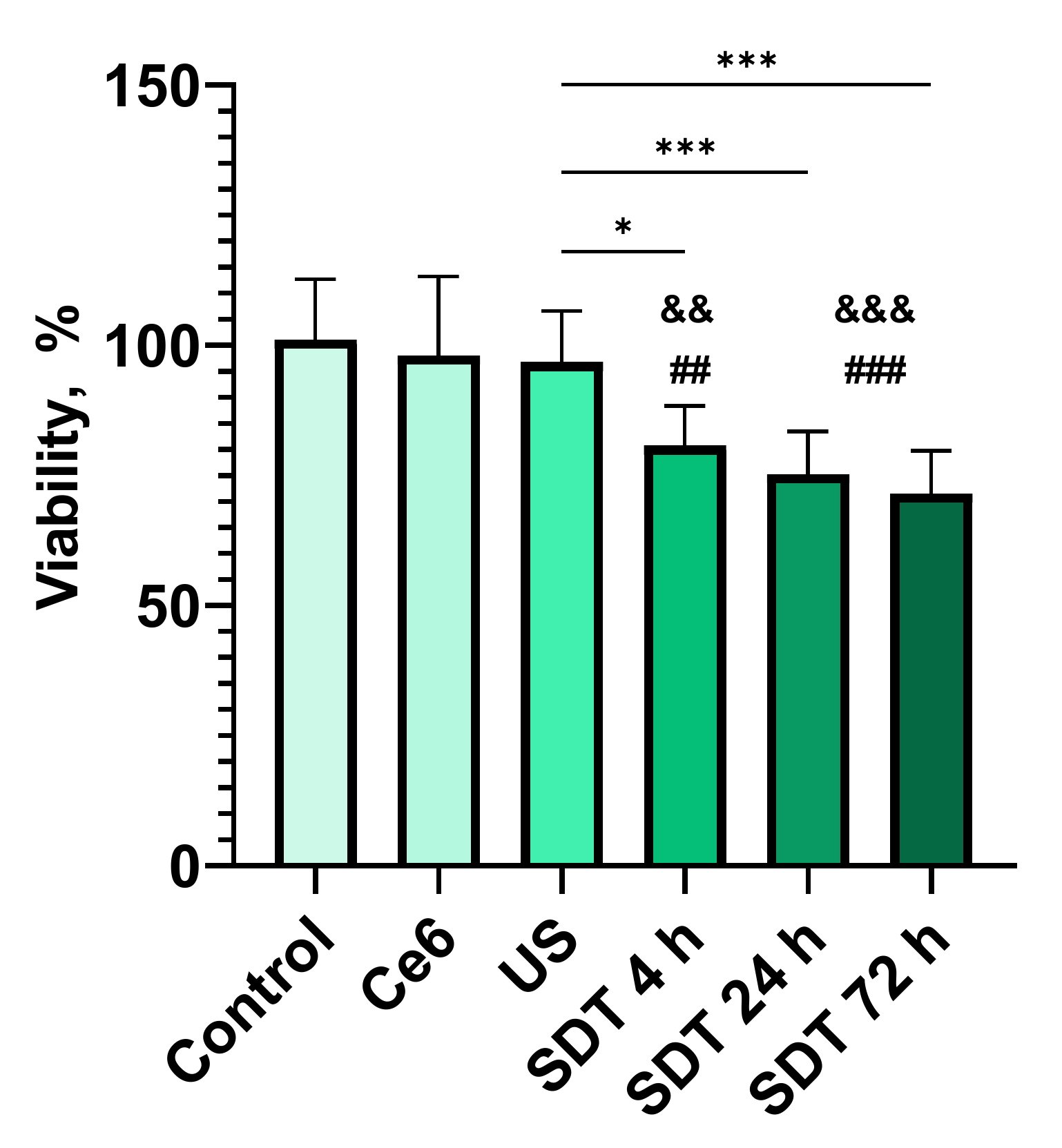}
	\caption{Viability of A375 cells incubated with Ce6 (10$^{-6}$ M) for 4, 24 and 72 hours after exposure to 20 kHz US with intensity 0.083 W/cm$^2$ for 10 seconds. The viability of cells, cells incubated with Ce6 and cells exposed to US were used as reference samples. *p$<$0.05 SDT 4h group vs. US group; ***p$<$0.001 SDT 24h/SDT 72h group vs. US group; $^{\sharp \sharp}$p$<$0.01 SDT 4h group vs. Control group; $^{\sharp\sharp \sharp}$p$<$0.001 SDT 24h/SDT 72h group vs. Control group; $^{\&\&}$p$<$0.01 SDT 4h group vs. Ce6 group; $^{\&\&\&}$p$<$0.001 SDT 24h/SDT 72h group vs. Ce6 group} 
	\label{fgr:1}
\end{figure}

Figure \ref{fgr:1} shows the viability of A375 cells incubated with Ce6 ($10^{-6}$ M) for 4, 24 and 72 hours after exposure to 20 kHz US with power 0.8 W for 10 seconds (for details see the section SDT test at Materials and Methods). The US energy incident on the sample was calculated as 8 J. According to data presented in Figure 1, exposure of 20 kHz US on A375 cells incubated with Ce6 leads to more than 25$\%$ cells death, demonstrating for the first time that combining Ce6 with the low-frequency US can produce SDT effect. The works presented in  literature have been discussing the causes of the \cite{pan2018metal, liu2017multifunctional} sonodynamic effect only as a result of exposure to therapeutic US (1 – 3 MHz, 0.5 – 3 W/cm$^2$).

\newpage
Figure 1 clearly demonstrates that Ce6 does not show any cytotoxicity effect on A375 melanoma cells. At the same, time the situation changes when cells incubated with the same amount of Ce6 were exposed to US. The US treatment of A375 melanoma cells with Ce6 molecules results in the death of the cell, while the US irradiation only on A375 melanoma cells shows no effect. It is of interest to find out the mechanism that is responsible for the appearance of the sonodynamic effect. Previously, we discussed two mechanisms that can lead to the destruction of cancer cells \cite{nikolaev2009use, wan2016recent}. The first mechanism is cavitation and the use of aggregates of Ce6 molecules to enhance the cavitation effect that leads to the destruction of cancerous \cite{paris2018ultrasound, shanei2020effect}.  This mechanism has not been demonstrated for molecules in monomeric form. The presence of aggregates of Ce6 molecules can enhance this cavitation effect, leading to some thermal or chemical local effects that destroy the cells due to cavitation. In this case, the aggregates of molecules are used as centers of cavitation \cite{nikolaev2015combined}. The second mechanism is based on SO generation by the Ce6 molecules under US exposure. It requires the presence of molecules in monomeric form because their aggregates are usually less efficient SO generator \cite{paul2013elucidation}. 

Therefore, the main SDT mechanism is not clear and can be elucidated relying on the Ce6 intracellular form (monomers or aggregates) in cancer cells. The presence of monomers or aggregated molecules can be assessed by the traditional PL and PL excitation spectra \cite{yu1995experimental}. If Ce6 molecules take the aggregate form, the main sonodynamic mechanism is cavitation. If Ce6 molecules take the monomeric form, the main sonodynamic mechanism is Ce6 molecules SO generation by sonoluminescence (SL).

\subsection{Chlorin e6 optical properties in melanoma cells }
PDT effect of tetrapyrrole molecules strongly depends on their SO generation efficiency. It is well known that aggregation of tetrapyrrole molecules leads to the loss of their ability to generate SO because of the increase of non-radiative rate in tetrapyrrole aggregates \cite{paul2016ph, parkhats2009dynamics}. Ce6 is a tetrapyrrole molecule and its optical properties are well studied in water solution at different pH levels, dimethyl sulfoxide (DMSO) and PBS \cite{mojzisova2007ph, paul2013optimization}. At the same time, there are no data on absorption and photoluminescence (PL) spectra of Ce6 in any cancer cells. Therefore, we have examined PL and PL excitation spectra of Ce6 in A375 melanoma cells. According to the spectra, we can evaluate the state of our molecules directly in the cancer cells. All measurements were performed in 12-well plates (for details see the section Spectroscopic measurements at Materials and Methods). 
Figure \ref{fgr:2a, 2b} shows the PL and PL excitation spectra of Ce6 in a RPMI nutrient medium and inside the cancer cells.

\begin{figure}[h]
	\centering
	\includegraphics[height=4.5 cm ]{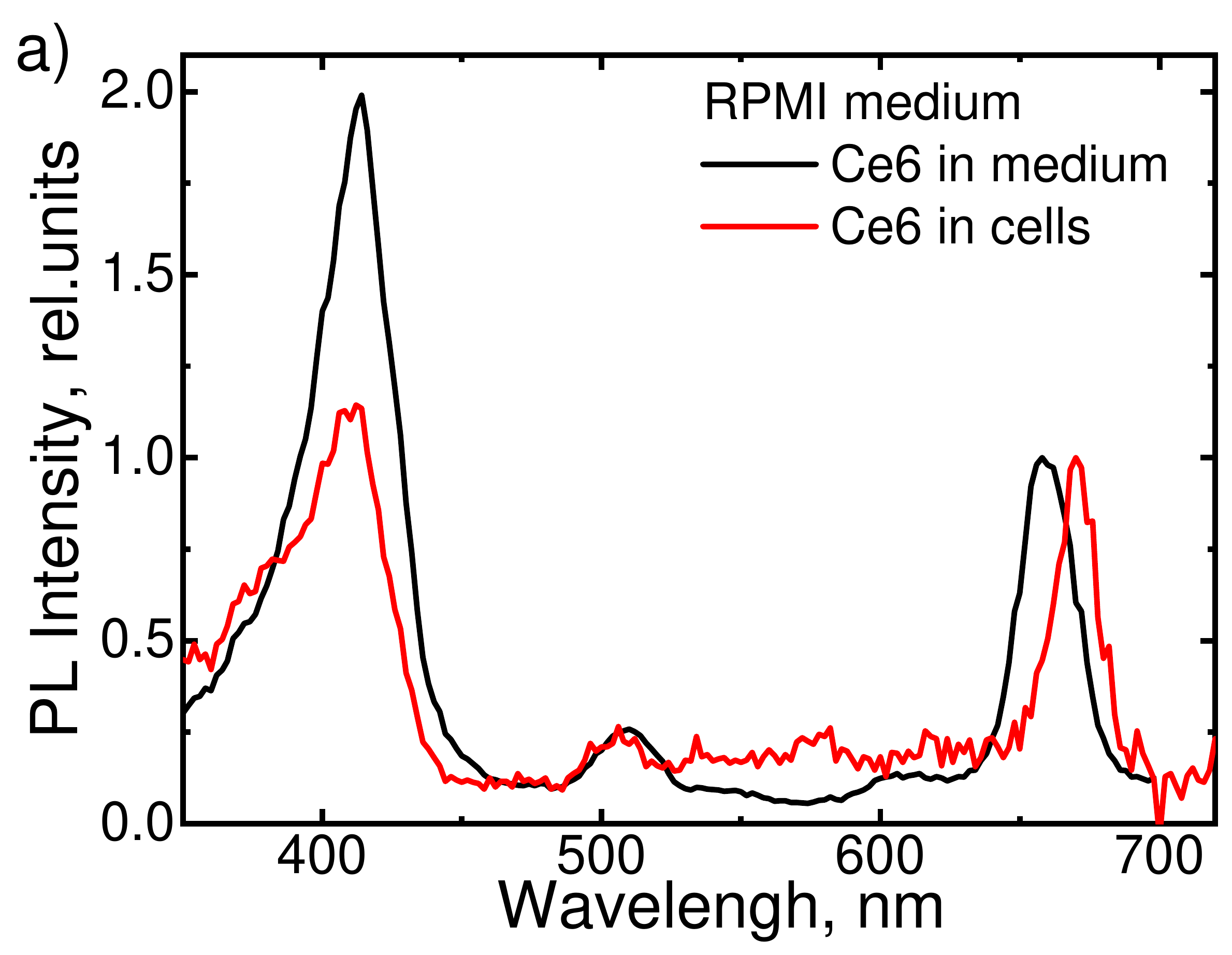} \hfill
		\includegraphics[height=4.5 cm ]{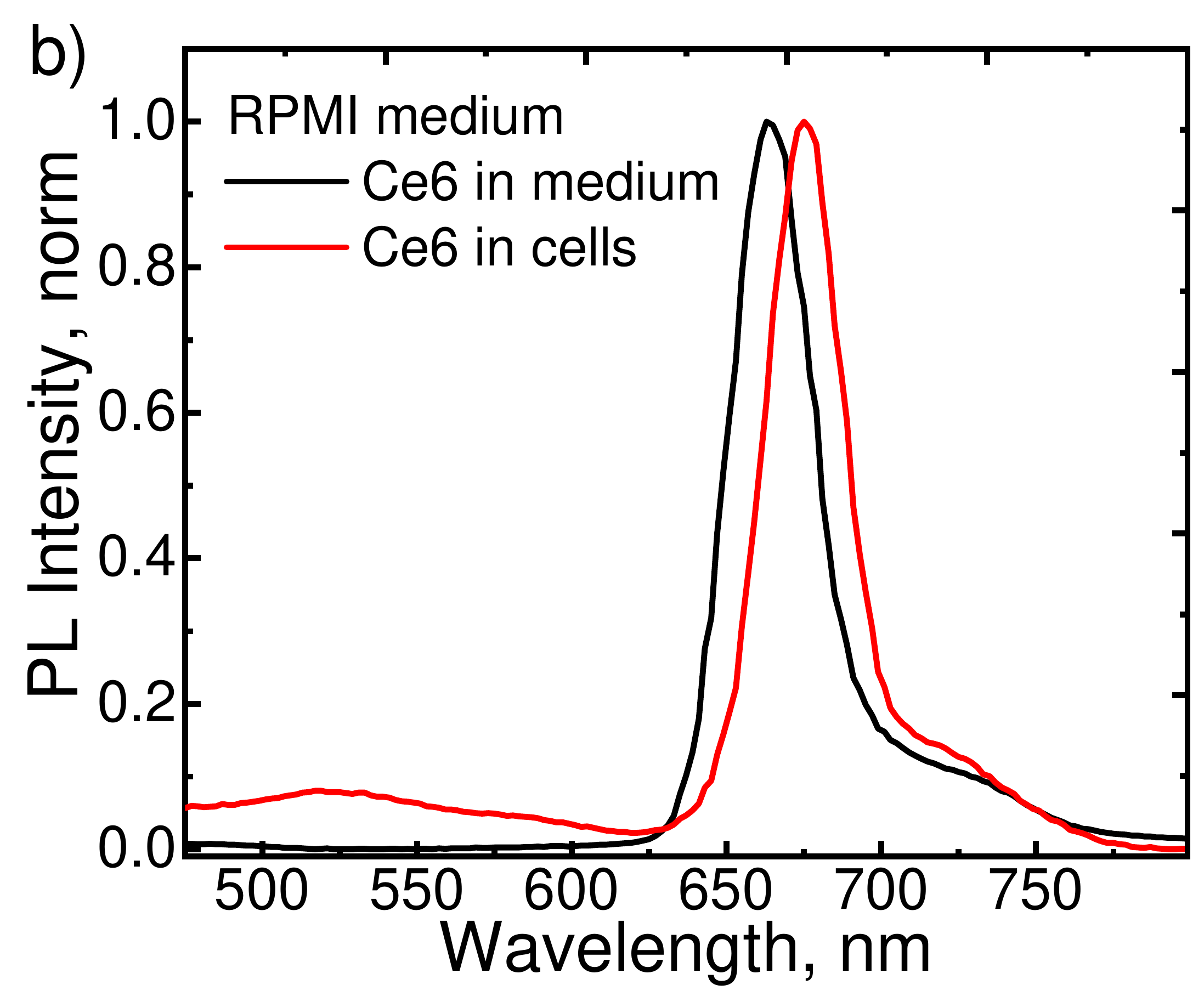}

	\caption{(a) PL excitation spectra of Ce6 in bare RPMI medium (black curve) and Ce6 in cells (red curve). PL excitation spectra are normalized by maximum of first absorption band of Ce6. (b) Normalized PL spectra of Ce6 in bare RPMI medium (black curve) and Ce6 in cells with RPMI (red curve). PL excitation wavelength is 405 nm, PL registration wavelength is 760 nm.}
	\label{fgr:2a, 2b}
\end{figure}

\newpage
Figure \ref{fgr:2a, 2b} shows the Ce6 PL excitation and PL spectra, which confirms the monomeric form of Ce6, because of the position of the Ce6 PL band at 676 nm, the Q(IV) absorption band at 500 nm, and the position of the Q(I) band at 670 nm. It should also be noted that the Ce6 monomer in A375 cells demonstrates the “reddest” position of its PL band in comparison with RPMI medium and PBS solutions, representing a sample in a strong alkaline environment that is not characteristic of cancer cells \cite{gerweck1996cellular}. We believe that Ce6 accumulates in mitochondria \cite{li2014effects, wang2015anti}, and this allows the Ce6 to remain in its monomeric form.
Also, in the section A of Supplementary materials, the stability of Ce6 monomers was shown under the action of 20 kHz US. Consequently, US of this frequency and power, does not lead to Ce6 aggregation. According to the results, Ce6 molecules do not aggregate in cancer cells with low pH or when exposed to US. It was previously shown (in Figure \ref{fgr:1}) that US at this intensity and frequency itself does not lead to the destruction of cancer cells, for which also Ce6 molecules are needed. It means that the SDT mechanism due to the formation of a cavitation nucleus and Ce6 aggregation is unlikely in our case.

\subsection{SDT vs PDT efficiency in melanoma cells incubated with Ce6 molecules}
\begin{figure}[h]
	\centering
	\includegraphics[height=6 cm ]{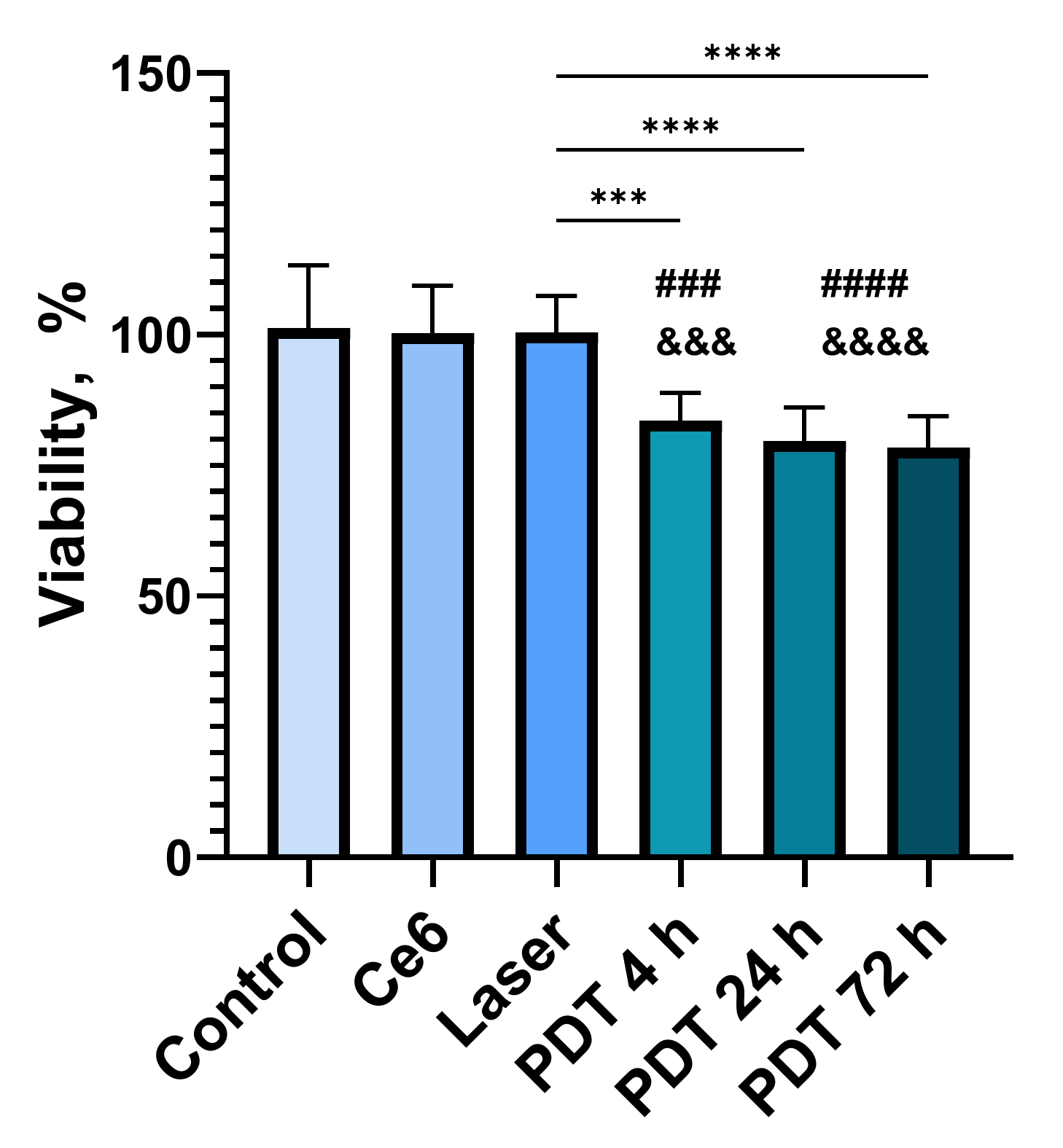}
	\caption{Viability of A375 cells incubated with Ce6 ($10^{-6}$ M) for 4, 24, 72 hours after their irradiation with 650 nm LED for 117 sec. The viability of cells, cells incubated with Ce6 and cells exposed to LED were used as reference samples. ***p$<$0.001 PDT 4h group vs. laser group; ****p$<$0.0001 PDT 24h/PDT 72h group vs. laser group; $^{\sharp\sharp\sharp}$p$<$0.001 PDT 4h group vs. Control group; $^{\sharp\sharp\sharp\sharp}$p$<$0.0001 PDT 24h/PDT 72h group vs. Control group; $^{\&\&\&}$p$<$0.001 PDT 4h group vs. Ce6 group; $^{\&\&\&\&}$p$<$0.0001 PDT 24h/PDT 72h group vs. Ce6 group} 
	\label{fgr:3}
\end{figure}

Considering the second SDT mechanism, which leads to the generation of SO, by the combined action of US and photosensitizer molecules. In this model the SDT effect is due to the SO generation by photosensitizers, which is a two-step  process. At the first step, US produces cavitation effect in cells’ medium or in cells, that can produce SL characterized with a wide band in visible spectral range (400-1100 nm) \cite{kozabaranov2020piezoceramic}. At the next step, this light can be absorbed by photosensitizer, i.e. Ce6 in our case, transferring it to $T_{1}$ state \cite{derosa2002photosensitized}. Thus, US is used to obtain SL, which serves as a source, similar to visible light. It is well documented that US at a frequency of 20 kHz is capable to generate a SL \cite{suslick2008inside}. 
We assume that SL is formed under the action of US, then the efficiency of the photodynamic effect and the sonodynamic effect at the same energies should be similar. Comparison between the impact of US and light on A375 cells loaded with Ce6 is shown below (Figure \ref{fgr:3}).
The photodynamic test is used as a reference experiment to confirm that Ce6 can efficiently generate SO in cancer cells. The viability of cells incubated with Ce6 under exposed by light is shown in Figure \ref{fgr:3}.
\newpage

\begin{figure}[h]
	\centering
	\includegraphics[height=6 cm ]{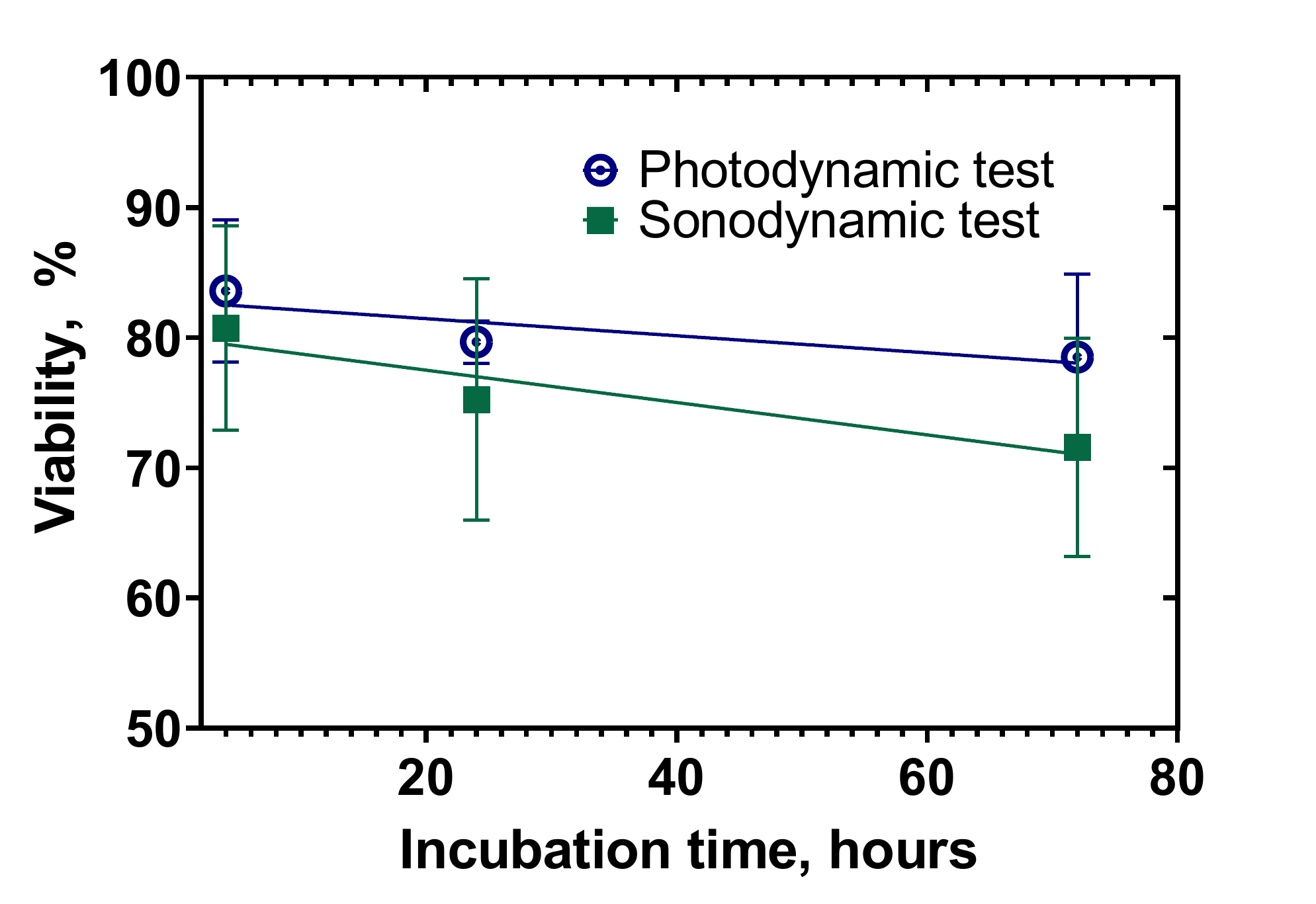}
	\caption{The dependence of cell viability after sonodynamic test (green curve) and photodynamic test (blue curve) on the incubation time of A375 cells with Ce6 molecules with the same incident energy on the samples (8 J). Parameters of US: 20 kHz with power 0.8 W for 10 seconds. Parameters for LED: 650 nm with power 20 mW for 117 seconds}
	\label{fgr:4}
\end{figure}

Viability data presented in Figure \ref{fgr:3} confirms PDT effect on A375 cells that has led to 19 \%,   22 \% and 23 \% decrease in cells viability with incubation time of 4, 24 and 72 hours, respectively. As clearly seen in Figure 3, Ce6 does not demonstrate any toxic effect on A375 cells after 24 hours of incubation. At the same time, the situation is dramatically changed when cells incubated with the same amount of Ce6 are irradiated by 650 nm LED. It is not surprising effect because Ce6 is a second-generation PDT drug that is widely used in the clinic \cite{ivanov2000one}. Thus, it becomes clear that under the influence of light, Ce6 behaves as a traditional photosensitizer. We suggest that the increased time of incubation enhanced the Ce6 intracellular concentration, which cause the  increase in PDT effect.
Figure \ref{fgr:4} shows the dependence of PDT and SDT efficiency on the incubation time of cells with Ce6 and with the same incident energy per sample.
\newpage

Figure \ref{fgr:4} compares PDT and SDT tests with the same energy incident on the sample. According to the results, the efficiency of photodynamic and sonodynamic tests increases with the increase in incubation time of cells with Ce6. In addition, it is clearly seen that an increase in the incubation time leads to an increase in the difference between PDT and SDT effects. At the same time, exposure to US in combination with Ce6 leads to greater cell death rate than exposure to light with Ce6. It is the first demonstration of the SDT effect with these configurations and further research is needed to clarify the observed effect. If we assume the second mechanism of SDT is active, then the sonodynamic effect can be associated with the formation of SL microbubbles inside the nutrient medium \cite{tronson2002comparison}.  Probably, Ce6 excitation arose from chaotic microbubbles formation is more efficient than exposure to external light. SL can appear at the temperatures from 20°C and higher at the frequency 20 kHz \cite{price2004sonoluminescence}. In those conditions, SL will be observed at US power density of about  0.09 - 0.2 W/cm$^2$ \cite{tronson2012behavior}. The presented literary parameters are consistent with our experimental parameters, so we assume that SL can occur in our case. $K.Yasui$ showed the appearance of SL from microbubbles at an ultrasonic frequency of 20-30 kHz \cite{yasui2002influence}. In that work, the author calculated the energy of the light emitted during collapse of a single bubble in water to be is 7.2 pJ. At the same time, the energy of LED light absorbed by Ce6 molecules in cells equals 16 mJ (details are discussed in the Section C of Supplementary Materials). $M.Kazachek$ showed that at an US frequency of 20 kHz, about 10$^5$ sonoluminescent microbubbles can appear dirung one US period (50 $\mu$s) \cite{kazachek2020using}. An increase in the time of US exposure can lead to a greater number of sonoluminescent bubbles. Our results demonstrate that the combination of 20 kHz US with Ce6 leads to SDT effect probably due to the generation of SO by Ce6. We suggest that cavitation in the nutrient medium leads to the SL, which in its turn serves as a local internal light source for photoexcitation of Ce6 molecules.

\section{Conclusion}
Optical properties of Ce6 and effect of US and visible light have been studied. The impact of low-frequency US on the monomeric form of Ce6 molecules did not lead to structural changes in the Ce6 molecule. The combined effect of 20 kHz US and Ce6 on melanoma cells leads to a pronounced SDT effect. The mechanism of SDT in A375 melanoma cells is most likely based on photoexcitation of Ce6 in resulting the absorption of light by sonoluminescence. The sonoluminescence can originate from cavitation in the nutrient medium or cells. 

\section{Acknowledgments}
This work was supported by the Ministry of Science and Higher Education of the Russian Federation, GOSZADANIE No. 2019-1080.

\newpage

\appendix
\counterwithin{figure}{subsection}

\section*{Supplementary Materials}
 \addcontentsline{toc}{section}{Supplementary Materials}
 \renewcommand{\thesubsection}{\Alph{subsection}}

     \subsection{Optical properties of Ce6 after influence of US}
The absorption and PL spectra of Ce6 in DMSO before and after exposure to US presented in Figure \ref{fgr:A1, A2}. The exposure time for the samples is 30 seconds. The incident US energy is 27 J.

\begin{figure}[h]
	\centering
		\includegraphics[height=4.5 cm ]{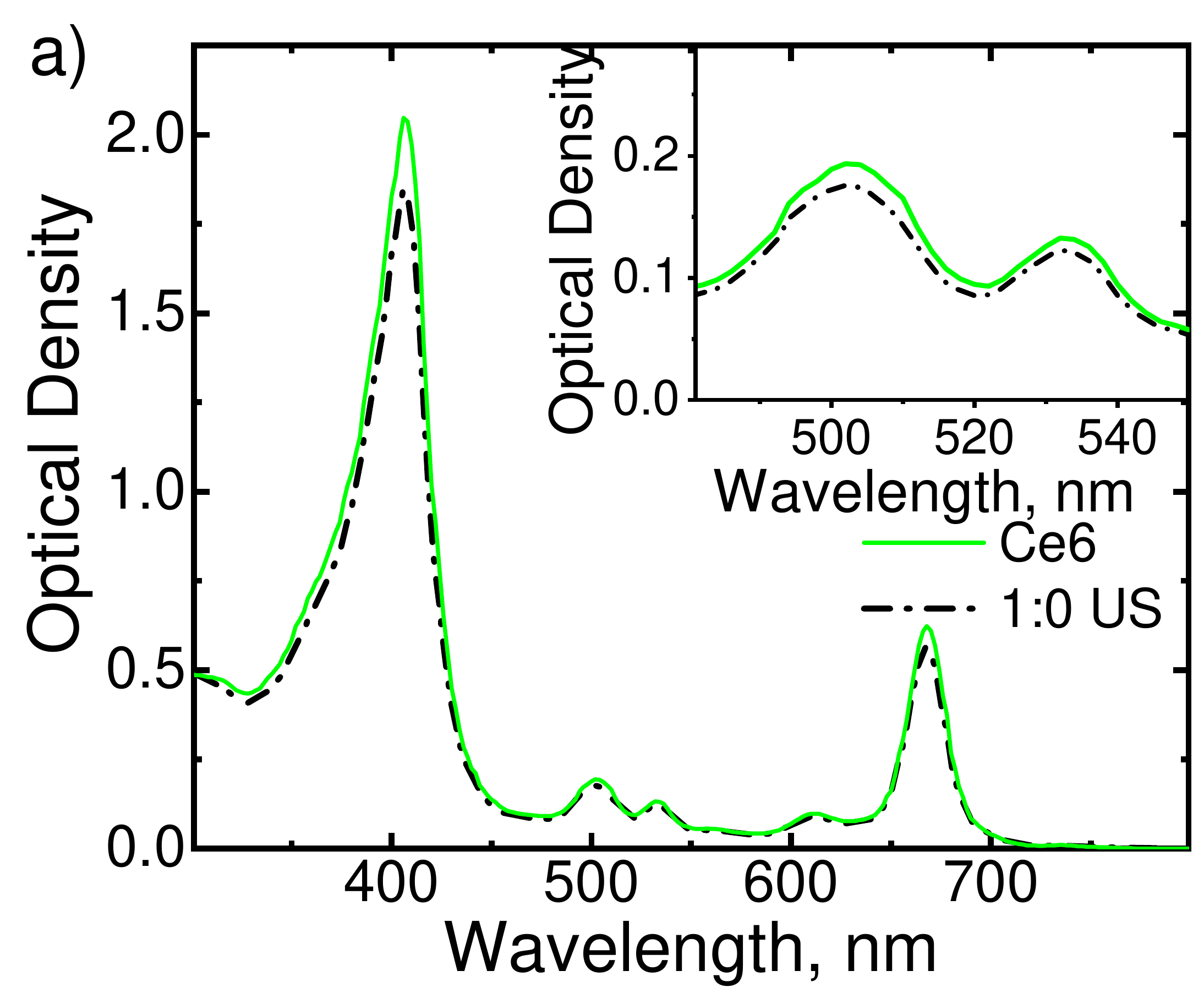} \hfill
		\includegraphics[height=4.5 cm ]{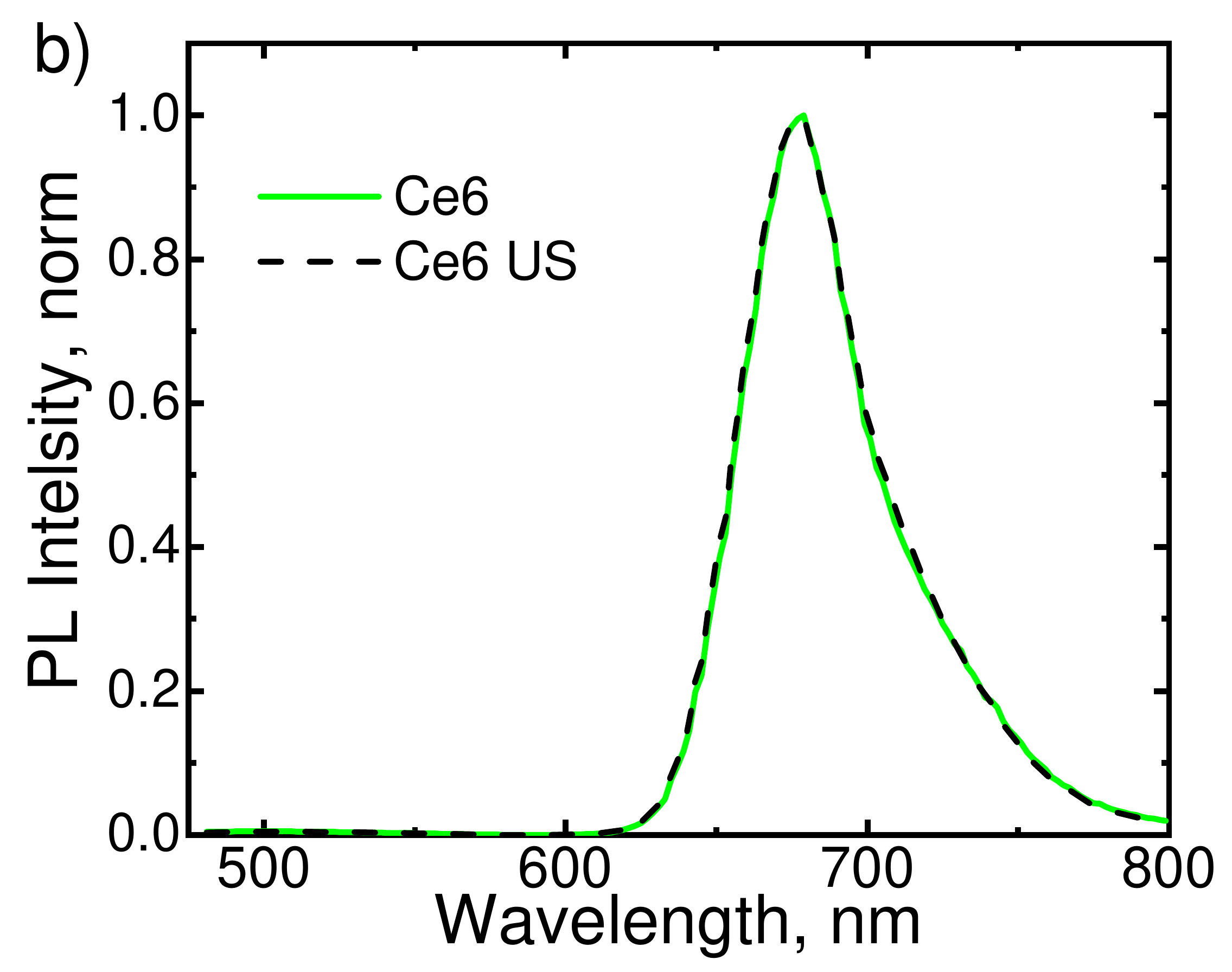}
	\caption{Absorption spectra (a) and normalized PL spectra (b) of Ce6 in DMSO before and after exposure to US; PL excitation wavelength is 435 nm}
	\label{fgr:A1, A2}
\end{figure}

An analysis of the absorption and PL spectra of Ce6 in a DMSO shows that the action of US at a frequency of 20 kHz does not lead to the dimerization \cite{visheratina2019circular} of the monomeric form of Ce6. At the same time, according to a decrease in the optical density in the band by 405 nm, the influence of low-frequency US of 20 kHz leads to a small drop in the optical density of Ce6. Thus, the concentration of Ce6 in the samples decreased, and this may indicate that under the action of US, SO is generated. As known, the appearance of SO can also lead to destruction of the Ce6 molecule \cite{derosa2002photosensitized}.

\subsection{Optical properties of Ce6 in a nutrient medium }
Figure \ref{fgr:B1_a, B2_b} demonstrates PL and PL excitation spectra of Ce6 in mixture of PBS:bidistilled water with the ratios 1:0 and 1:500.
\newpage

\begin{figure}[h]
	\centering
		\includegraphics[height=4.5 cm ]{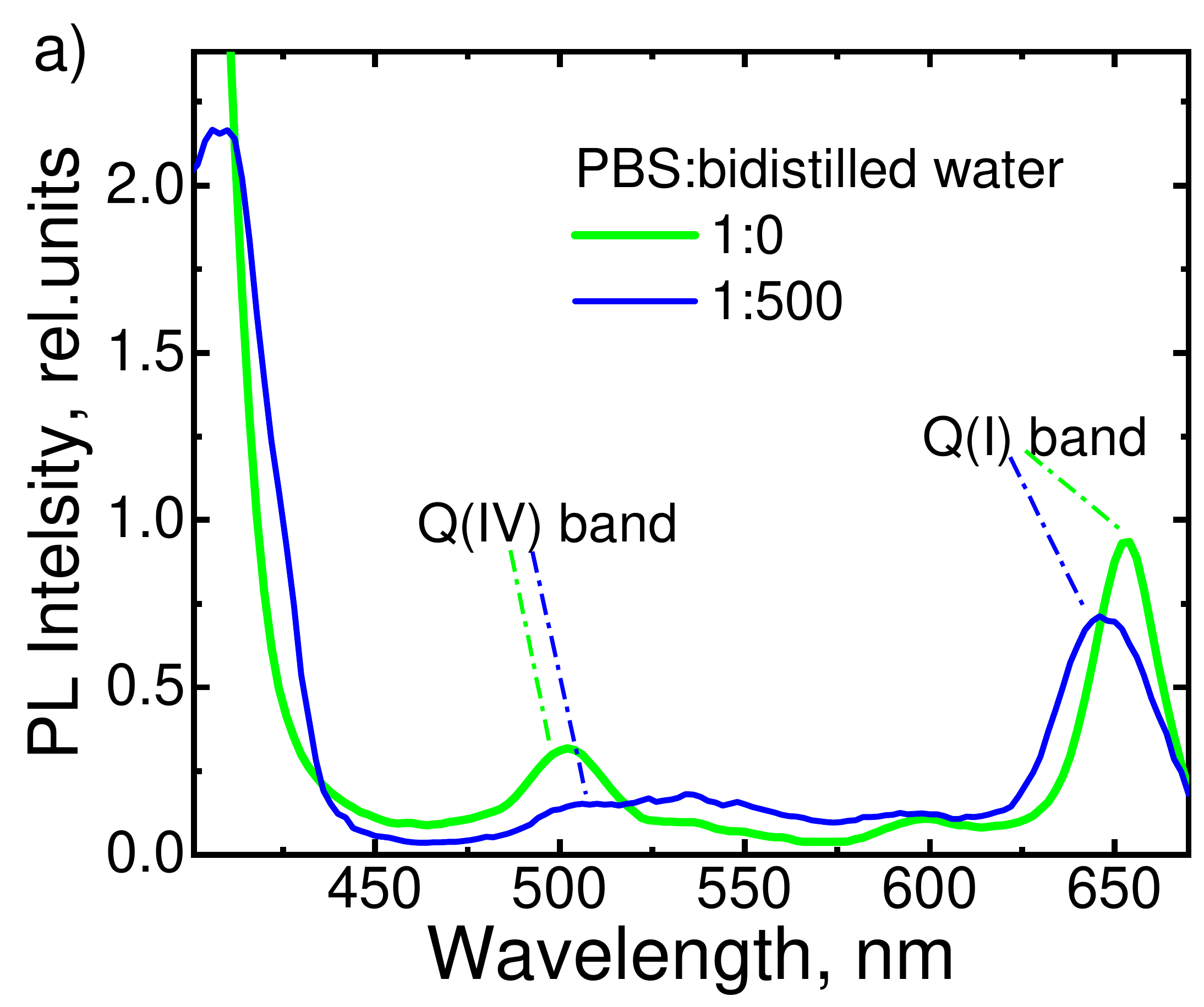} \hfill
		\includegraphics[height=4.5 cm ]{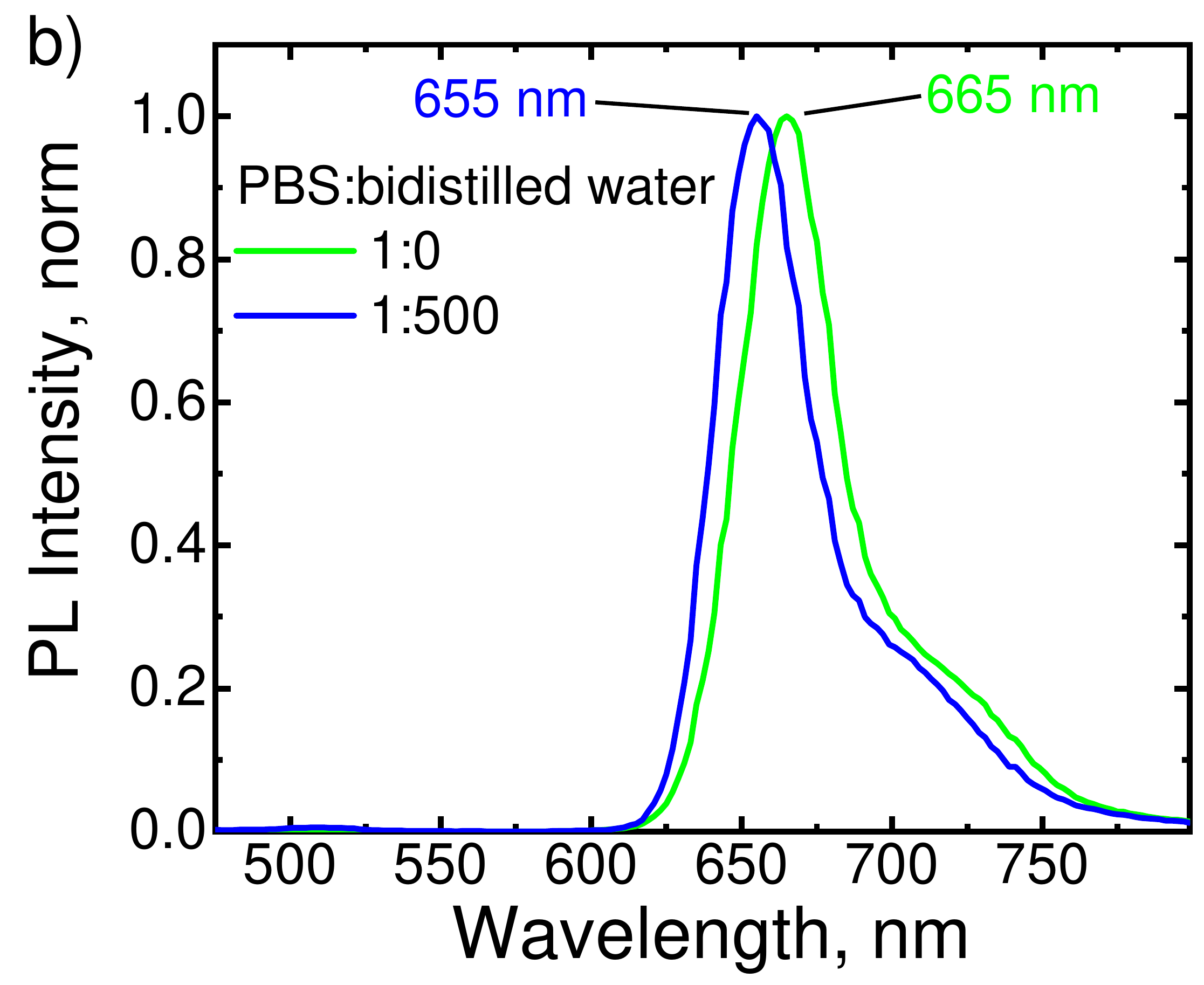}
	\caption{PL excitation (a) and PL (b) spectra of Ce6 in PBS and a PBS:bidistilled water mixture. Wavelengths at 435 nm and 700 nm were used as PL excitation and recording wavelength}
	\label{fgr:B1_a, B2_b}
\end{figure}

Analysis of Ce6 PL excitation and PL spectra presented in Figure \ref{fgr:B1_a, B2_b} showed that Ce6 in a 1:500 mixture is in the dimeric form. The dimeric form of Ce6 is confirms by the position of the several bands of Q(I) at 646 nm and Q(IV) at 500 nm, as well as PL band at 655 nm \cite{mojzisova2007ph}. Ce6 is a monomer in a 1:0 mixture, which is quite obvious since Ce6 has a strictly monomeric form in PBS \cite{jain2021fluorescence}.

Figure \ref{fgr:B1, B2} demonstrates PL and PL excitation spectra of Ce6 in a mixtures PBS:bidistilled water with the ratios 1:0 and 1:500 added to RPMI medium.

\begin{figure}[h]
	\centering
		\includegraphics[height=4.5 cm ]{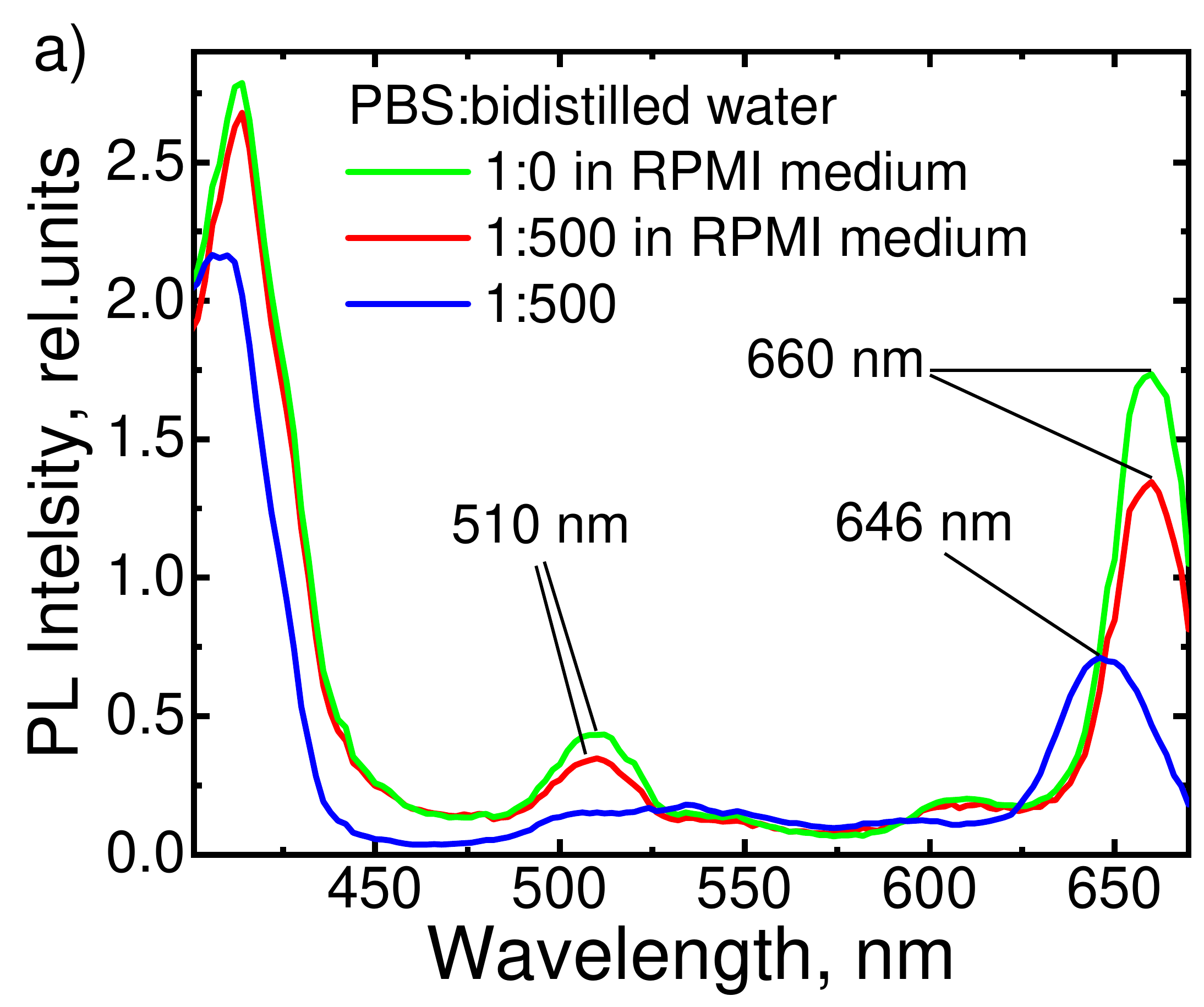} \hfill
		\includegraphics[height=4.5 cm ]{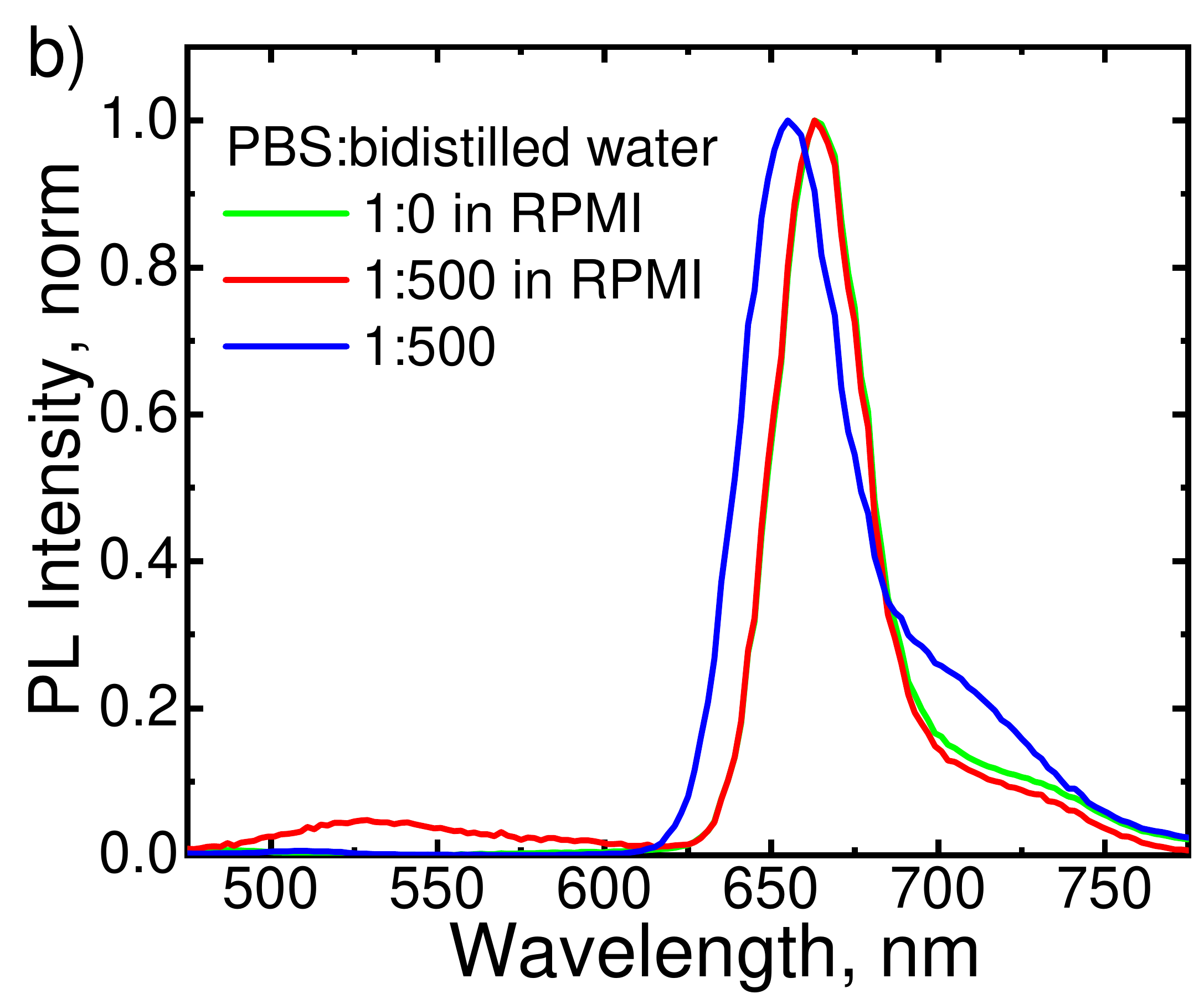}
	\caption{PL excitation (a) and PL (b) spectra of Ce6 in PBS and a PBS:bidistilled water mixture added to RPMI medium. Wavelengths at 435 nm and 700 nm were used as PL excitation and recording wavelength}
	\label{fgr:B1, B2}
\end{figure}

The analysis of the spectra presented in Figure \ref{fgr:B1, B2} has shown that Ce6 is monomer in RPMI medium in both cases, i.e. when it was added as monomer  (1:0) and as dimer (1:500). First of all, the monomeric form of Ce6 is characterized by the position Q(I), Q(IV) and the Soret bands. It means that the monomeric form of Ce6 is the most stable form of Ce6 in this medium. 

\subsection{Calculation of the absorbed energy dose by Ce6}
The absorbed energy can be found by the following formula:  
$$E_{abs}=I_{650}\times D_{650}\times ln(10)\times t, $$

\noindent where $I_{650}$ source power at emission wavelength, $D_{650}$ optical density of emission wavelength, $t$ is time of exposure to LED emmision. As a result, Ce6 in the cells absorbed 16 mJ.

\begin{figure}[h]
	\centering
	\includegraphics[height=6 cm ]{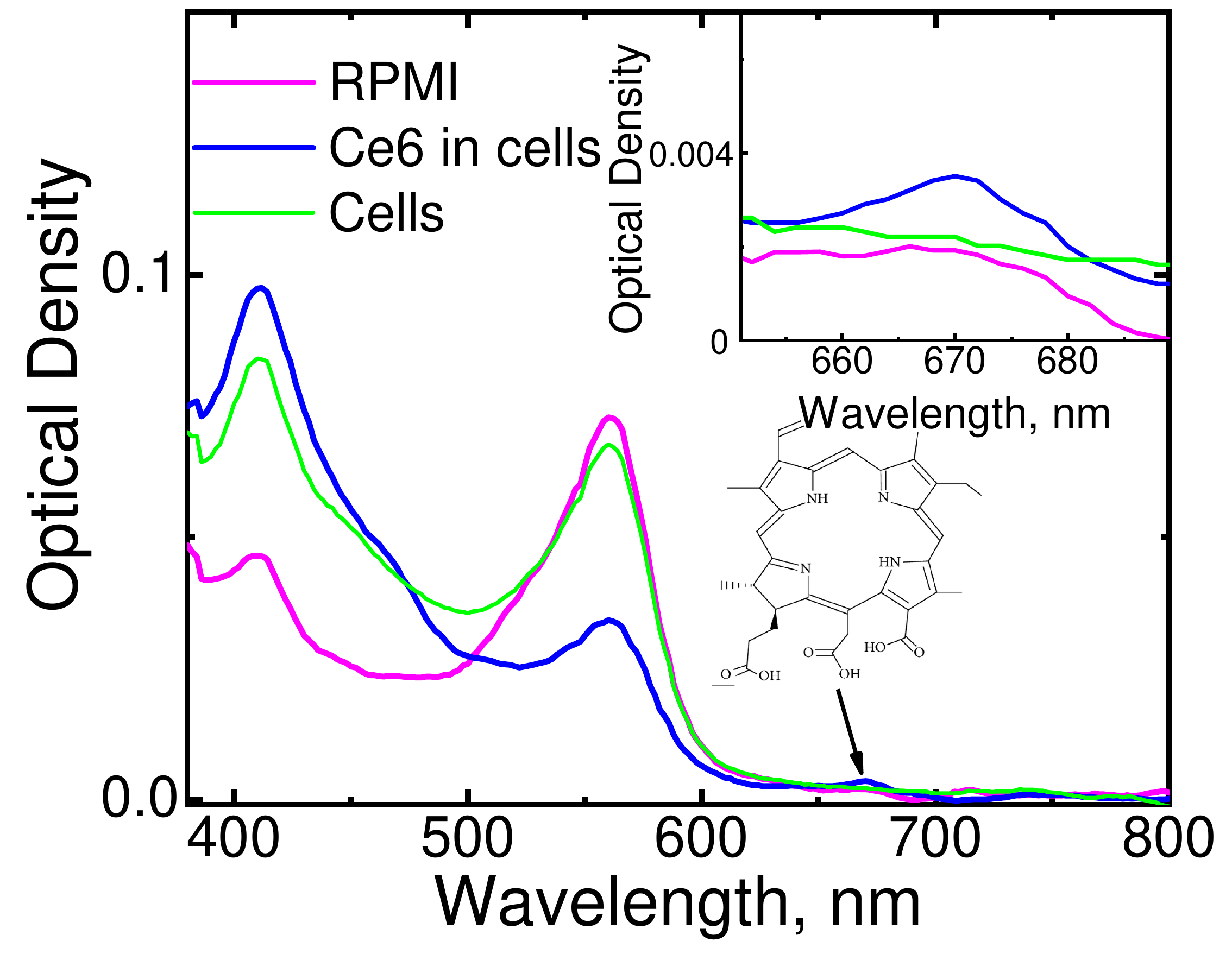}
	\caption{Absorption spectra of RPMI medium, cells and the cells, incubated with Ce6}
	\label{fgr:C}
\end{figure}

\newpage 

\subsection{Calculation of US power}
The intensity of US was estimated using the following formula:
$$P=\frac{E}{t\times A}   \quad   [\frac{W}{cm^2}],$$

\noindent where $t$ is the US exposure time, $E$ is incident US energy, A is the area of single well. The area of single well from 12-well plate is used as the surface area to evaluate the intensity of incident US wave directly on the sample. The US energy can be found by the following formula:

$$E=m\times C_{p}\times (T_{after}-T_{before}) \quad [J],$$

\noindent where $m$ is the mass of water in the single well, $C_{p}$ is the heat capacity of water, $T_{after}$ is the temperature of water after US exposure, ${T}_{before}$ is the temperature of water before US exposure.

Mass of the water can be found by the following formulas:

$$m=\rho\times V \quad [kg],$$
$$V=\pi\times r^2\times h \quad [m^3],$$
$$\rho=\frac{\rho_{before}+\rho_{after}}{2} \quad [\frac{kg}{m^3}],$$

\noindent where ${\rho_{before}}$ is the density of the water corresponding to temperature $T_{before}$, while $\rho_{after}$ is the density of water corresponding to temperature $T_{after}$.

\[{C_p} = \frac{{{C_{before}} + {C_{after}}}}{2}{\rm{ }}\left[ {\frac{J}{{kg\cdot K}}} \right],\]

\noindent where $C_{before}$ is the heat capacity of water corresponding to temperature $T_{before}$, while $C_{after}$ is the heat capacity of water corresponding to temperature $T_{after}$.

\bibliography{mybibfile}

\begin{thebibliography}{10}
\expandafter\ifx\csname url\endcsname\relax
  \def\url#1{\texttt{#1}}\fi
\expandafter\ifx\csname urlprefix\endcsname\relax\def\urlprefix{URL }\fi
\expandafter\ifx\csname href\endcsname\relax
  \def\href#1#2{#2} \def\path#1{#1}\fi

\bibitem{jones2007epigenomics}
P.~A. Jones, S.~B. Baylin, The epigenomics of cancer, Cell 128~(4) (2007)
  683--692.

\bibitem{abdel2016photodynamic}
M.~H. Abdel-Kader, Photodynamic therapy, Springer, 2016.

\bibitem{plaetzer2009photophysics}
K.~Plaetzer, B.~Krammer, J.~Berlanda, F.~Berr, T.~Kiesslich, Photophysics and
  photochemistry of photodynamic therapy: fundamental aspects, Lasers in
  medical science 24~(2) (2009) 259--268.

\bibitem{henderson2020photodynamic}
B.~W. Henderson, Photodynamic therapy: basic principles and clinical
  applications, CRC Press, 2020.

\bibitem{lan2019photosensitizers}
M.~Lan, S.~Zhao, W.~Liu, C.-S. Lee, W.~Zhang, P.~Wang, Photosensitizers for
  photodynamic therapy, Advanced healthcare materials 8~(13) (2019) 1900132.

\bibitem{dos2019photodynamic}
A.~F. dos Santos, D.~R.~Q. de~Almeida, L.~F. Terra, M.~S. Baptista,
  L.~Labriola, Photodynamic therapy in cancer treatment-an update review,
  Journal of Cancer Metastasis and Treatment 5.

\bibitem{yang2019reactive}
B.~Yang, Y.~Chen, J.~Shi, Reactive oxygen species (ros)-based nanomedicine,
  Chemical reviews 119~(8) (2019) 4881--4985.

\bibitem{yardeni2019host}
T.~Yardeni, C.~E. Tanes, K.~Bittinger, L.~M. Mattei, P.~M. Schaefer, L.~N.
  Singh, G.~D. Wu, D.~G. Murdock, D.~C. Wallace, Host mitochondria influence
  gut microbiome diversity: A role for ros, Science signaling 12~(588).

\bibitem{simon2000role}
H.-U. Simon, A.~Haj-Yehia, F.~Levi-Schaffer, Role of reactive oxygen species
  (ros) in apoptosis induction, Apoptosis 5~(5) (2000) 415--418.

\bibitem{fortes2012heme}
G.~B. Fortes, L.~S. Alves, R.~de~Oliveira, F.~F. Dutra, D.~Rodrigues, P.~L.
  Fernandez, T.~Souto-Padron, M.~J. De~Rosa, M.~Kelliher, D.~Golenbock, et~al.,
  Heme induces programmed necrosis on macrophages through autocrine tnf and ros
  production, Blood, The Journal of the American Society of Hematology 119~(10)
  (2012) 2368--2375.

\bibitem{taquet2007phthalocyanines}
J.-p. Taquet, C.~Frochot, V.~Manneville, M.~Barberi-Heyob, Phthalocyanines
  covalently bound to biomolecules for a targeted photodynamic therapy, Current
  medicinal chemistry 14~(15) (2007) 1673--1687.

\bibitem{agostinis2011photodynamic}
P.~Agostinis, K.~Berg, K.~A. Cengel, T.~H. Foster, A.~W. Girotti, S.~O.
  Gollnick, S.~M. Hahn, M.~R. Hamblin, A.~Juzeniene, D.~Kessel, et~al.,
  Photodynamic therapy of cancer: an update, CA: a cancer journal for
  clinicians 61~(4) (2011) 250--281.

\bibitem{mao2017chemiluminescence}
D.~Mao, W.~Wu, S.~Ji, C.~Chen, F.~Hu, D.~Kong, D.~Ding, B.~Liu,
  Chemiluminescence-guided cancer therapy using a chemiexcited photosensitizer,
  Chem 3~(6) (2017) 991--1007.

\bibitem{magalhaes2016chemiluminescence}
C.~M. Magalhaes, J.~C. Esteves~da Silva, L.~Pinto~da Silva, Chemiluminescence
  and bioluminescence as an excitation source in the photodynamic therapy of
  cancer: a critical review, ChemPhysChem 17~(15) (2016) 2286--2294.

\bibitem{you2016ros}
D.~G. You, V.~Deepagan, W.~Um, S.~Jeon, S.~Son, H.~Chang, H.~I. Yoon, Y.~W.
  Cho, M.~Swierczewska, S.~Lee, et~al., Ros-generating tio 2 nanoparticles for
  non-invasive sonodynamic therapy of cancer, Scientific reports 6~(1) (2016)
  1--12.

\bibitem{xu2016sonodynamic}
C.~Xu, J.~Dong, M.~Ip, X.~Wang, A.~W. Leung, Sonodynamic action of chlorin e6
  on staphylococcus aureus and escherichia coli, Ultrasonics 64 (2016) 54--57.

\bibitem{wang2011detection}
J.~Wang, Y.~Guo, B.~Liu, X.~Jin, L.~Liu, R.~Xu, Y.~Kong, B.~Wang, Detection and
  analysis of reactive oxygen species (ros) generated by nano-sized tio2 powder
  under ultrasonic irradiation and application in sonocatalytic degradation of
  organic dyes, Ultrasonics Sonochemistry 18~(1) (2011) 177--183.

\bibitem{yasui2018acoustic}
K.~Yasui, Acoustic cavitation and bubble dynamics, Springer, 2018.

\bibitem{tuziuti2005correlation}
T.~Tuziuti, K.~Yasui, M.~Sivakumar, Y.~Iida, N.~Miyoshi, Correlation between
  acoustic cavitation noise and yield enhancement of sonochemical reaction by
  particle addition, The Journal of Physical Chemistry A 109~(21) (2005)
  4869--4872.

\bibitem{tuziuti2004effect}
T.~Tuziuti, K.~Yasui, Y.~Iida, H.~Taoda, S.~Koda, Effect of particle addition
  on sonochemical reaction, Ultrasonics 42~(1-9) (2004) 597--601.

\bibitem{canaparo2020bright}
R.~Canaparo, F.~Foglietta, F.~Giuntini, A.~Francovich, L.~Serpe, The bright
  side of sound: perspectives on the biomedical application of
  sonoluminescence, Photochemical \& Photobiological Sciences 19~(9) (2020)
  1114--1121.

\bibitem{ohmura2011sonodynamic}
T.~Ohmura, T.~Fukushima, H.~Shibaguchi, S.~Yoshizawa, T.~Inoue, M.~Kuroki,
  K.~Sasaki, S.-I. Umemura, Sonodynamic therapy with 5-aminolevulinic acid and
  focused ultrasound for deep-seated intracranial glioma in rat, Anticancer
  research 31~(7) (2011) 2527--2533.

\bibitem{beguin2019direct}
E.~Beguin, S.~Shrivastava, N.~V. Dezhkunov, A.~P. McHale, J.~F. Callan,
  E.~Stride, Direct evidence of multibubble sonoluminescence using therapeutic
  ultrasound and microbubbles, ACS applied materials \& interfaces 11~(22)
  (2019) 19913--19919.

\bibitem{giuntini2018insight}
F.~Giuntini, F.~Foglietta, A.~M. Marucco, A.~Troia, N.~V. Dezhkunov,
  A.~Pozzoli, G.~Durando, I.~Fenoglio, L.~Serpe, R.~Canaparo, Insight into
  ultrasound-mediated reactive oxygen species generation by various
  metal-porphyrin complexes, Free Radical Biology and Medicine 121 (2018)
  190--201.

\bibitem{miller2012overview}
D.~L. Miller, N.~B. Smith, M.~R. Bailey, G.~J. Czarnota, K.~Hynynen, I.~R.~S.
  Makin, B.~C. of~the American Institute of Ultrasound~in Medicine, Overview of
  therapeutic ultrasound applications and safety considerations, Journal of
  ultrasound in medicine 31~(4) (2012) 623--634.

\bibitem{park2014sonophoresis}
D.~Park, H.~Park, J.~Seo, S.~Lee, Sonophoresis in transdermal drug deliverys,
  Ultrasonics 54~(1) (2014) 56--65.

\bibitem{katz2004rapid}
N.~P. Katz, D.~E. Shapiro, T.~E. Herrmann, J.~Kost, L.~M. Custer, Rapid onset
  of cutaneous anesthesia with emla cream after pretreatment with a new
  ultrasound-emitting device, Anesthesia \& Analgesia 98~(2) (2004) 371--376.

\bibitem{mitragotri2004low}
S.~Mitragotri, J.~Kost, Low-frequency sonophoresis: a review, Advanced drug
  delivery reviews 56~(5) (2004) 589--601.

\bibitem{bhatnagar2016exploitation}
S.~Bhatnagar, J.~J. Kwan, A.~R. Shah, C.-C. Coussios, R.~C. Carlisle,
  Exploitation of sub-micron cavitation nuclei to enhance ultrasound-mediated
  transdermal transport and penetration of vaccines, Journal of Controlled
  Release 238 (2016) 22--30.

\bibitem{seto2010effects}
J.~E. Seto, B.~E. Polat, R.~F. Lopez, D.~Blankschtein, R.~Langer, Effects of
  ultrasound and sodium lauryl sulfate on the transdermal delivery of
  hydrophilic permeants: Comparative in vitro studies with full-thickness and
  split-thickness pig and human skin, Journal of Controlled Release 145~(1)
  (2010) 26--32.

\bibitem{mitragotri2006transdermal}
S.~Mitragotri, Transdermal drug delivery using low-frequency sonophoresis, in:
  BioMEMS and biomedical nanotechnology, Springer, 2006, pp. 223--236.

\bibitem{tezel2004topical}
A.~Tezel, S.~Dokka, S.~Kelly, G.~E. Hardee, S.~Mitragotri, Topical delivery of
  anti-sense oligonucleotides using low-frequency sonophoresis, Pharmaceutical
  research 21~(12) (2004) 2219--2225.

\bibitem{didenko2000multibubble}
Y.~T. Didenko, T.~Gordeychuk, Multibubble sonoluminescence spectra of water
  which resemble single-bubble sonoluminescence, Physical review letters
  84~(24) (2000) 5640.

\bibitem{ji2018multibubble}
R.~Ji, R.~Pflieger, M.~Virot, S.~I. Nikitenko, Multibubble sonochemistry and
  sonoluminescence at 100 khz: the missing link between low-and high-frequency
  ultrasound, The Journal of Physical Chemistry B 122~(27) (2018) 6989--6994.

\bibitem{edge2018singlet}
R.~Edge, T.~G. Truscott, Singlet oxygen and free radical reactions of retinoids
  and carotenoids—a review, Antioxidants 7~(1) (2018) 5.

\bibitem{bashkatov2005optical}
A.~Bashkatov, E.~Genina, V.~Kochubey, V.~Tuchin, Optical properties of human
  skin, subcutaneous and mucous tissues in the wavelength range from 400 to
  2000 nm, Journal of Physics D: Applied Physics 38~(15) (2005) 2543.

\bibitem{bharathiraja2017chlorin}
S.~Bharathiraja, M.~S. Moorthy, P.~Manivasagan, H.~Seo, K.~D. Lee, J.~Oh,
  Chlorin e6 conjugated silica nanoparticles for targeted and effective
  photodynamic therapy, Photodiagnosis and photodynamic therapy 19 (2017)
  212--220.

\bibitem{paszko2011nanodrug}
E.~Paszko, C.~Ehrhardt, M.~O. Senge, D.~P. Kelleher, J.~V. Reynolds, Nanodrug
  applications in photodynamic therapy, Photodiagnosis and photodynamic therapy
  8~(1) (2011) 14--29.

\bibitem{chin2009vivo}
W.~W. Chin, P.~S. Thong, R.~Bhuvaneswari, K.~C. Soo, P.~W. Heng, M.~Olivo,
  In-vivo optical detection of cancer using chlorin e6--polyvinylpyrrolidone
  induced fluorescence imaging and spectroscopy, BMC medical imaging 9~(1)
  (2009) 1--13.

\bibitem{pan2018metal}
X.~Pan, L.~Bai, H.~Wang, Q.~Wu, H.~Wang, S.~Liu, B.~Xu, X.~Shi, H.~Liu,
  Metal--organic-framework-derived carbon nanostructure augmented sonodynamic
  cancer therapy, Advanced materials 30~(23) (2018) 1800180.

\bibitem{liu2017multifunctional}
Y.~Liu, G.~Wan, H.~Guo, Y.~Liu, P.~Zhou, H.~Wang, D.~Wang, S.~Zhang, Y.~Wang,
  N.~Zhang, A multifunctional nanoparticle system combines sonodynamic therapy
  and chemotherapy to treat hepatocellular carcinoma, Nano Research 10~(3)
  (2017) 834--855.

\bibitem{nikolaev2009use}
A.~Nikolaev, A.~Gopin, V.~Bozhevol’nov, E.~Treshchalina, N.~Andronova,
  I.~Melikhov, Use of solid-phase inhomogeneities to increase the efficiency of
  ultrasonic therapy of oncological diseases, Acoustical Physics 55~(4) (2009)
  575--583.

\bibitem{wan2016recent}
G.-Y. Wan, Y.~Liu, B.-W. Chen, Y.-Y. Liu, Y.-S. Wang, N.~Zhang, Recent advances
  of sonodynamic therapy in cancer treatment, Cancer biology \& medicine 13~(3)
  (2016) 325.

\bibitem{paris2018ultrasound}
J.~L. Paris, C.~Mannaris, M.~V. Caba{\~n}as, R.~Carlisle, M.~Manzano,
  M.~Vallet-Reg{\'\i}, C.~C. Coussios, Ultrasound-mediated cavitation-enhanced
  extravasation of mesoporous silica nanoparticles for controlled-release drug
  delivery, Chemical Engineering Journal 340 (2018) 2--8.

\bibitem{shanei2020effect}
A.~Shanei, H.~Akbari-Zadeh, N.~Attaran, M.~R. Salamat,
  M.~Baradaran-Ghahfarokhi, Effect of targeted gold nanoparticles size on
  acoustic cavitation: An in vitro study on melanoma cells, Ultrasonics 102
  (2020) 106061.

\bibitem{nikolaev2015combined}
A.~Nikolaev, A.~Gopin, V.~Bozhevol’nov, H.~Treshalina, N.~Andronova,
  I.~Melikhov, D.~Filonenko, S.~Mazina, G.~Gerasimova, E.~Khorosheva, et~al.,
  Combined method of ultrasound therapy of oncological diseases, Russian
  Journal of General Chemistry 85~(1) (2015) 303--320.

\bibitem{paul2013elucidation}
S.~Paul, S.~Selvam, P.~W.~S. Heng, L.~W. Chan, Elucidation of monomerization
  effect of pvp on chlorin e6 aggregates by spectroscopic, chemometric,
  thermodynamic and molecular simulation studies, Journal of fluorescence
  23~(5) (2013) 1065--1076.

\bibitem{yu1995experimental}
J.~Yu, J.~Hsu, K.~Chuang, C.~Chao, S.~Chen, F.~Kao, W.~Fann, S.~Lin,
  Experimental and theoretical studies of absorption and photoluminescence
  excitation spectra of poly (p-phenylene vinylene), Synthetic metals 74~(1)
  (1995) 7--13.

\bibitem{paul2016ph}
S.~Paul, P.~W.~S. Heng, L.~W. Chan, ph-dependent complexation of
  hydroxypropyl-beta-cyclodextrin with chlorin e6: effect on solubility and
  aggregation in relation to photodynamic efficacy, Journal of Pharmacy and
  Pharmacology 68~(4) (2016) 439--449.

\bibitem{parkhats2009dynamics}
M.~Parkhats, V.~Galievsky, A.~Stashevsky, T.~Trukhacheva, B.~Dzhagarov,
  Dynamics and efficiency of the photosensitized singlet oxygen formation by
  chlorin e 6: The effects of the solution ph and polyvinylpyrrolidone, Optics
  and Spectroscopy 107~(6) (2009) 974--980.

\bibitem{mojzisova2007ph}
H.~Mojzisova, S.~Bonneau, C.~Vever-Bizet, D.~Brault, The ph-dependent
  distribution of the photosensitizer chlorin e6 among plasma proteins and
  membranes: a physico-chemical approach, Biochimica et Biophysica Acta
  (BBA)-Biomembranes 1768~(2) (2007) 366--374.

\bibitem{paul2013optimization}
S.~Paul, P.~W.~S. Heng, L.~W. Chan, Optimization in solvent selection for
  chlorin e6 in photodynamic therapy, Journal of fluorescence 23~(2) (2013)
  283--291.

\bibitem{gerweck1996cellular}
L.~E. Gerweck, K.~Seetharaman, Cellular ph gradient in tumor versus normal
  tissue: potential exploitation for the treatment of cancer, Cancer research
  56~(6) (1996) 1194--1198.

\bibitem{li2014effects}
Q.~Li, Q.~Liu, P.~Wang, X.~Feng, H.~Wang, X.~Wang, The effects of ce6-mediated
  sono-photodynamic therapy on cell migration, apoptosis and autophagy in mouse
  mammary 4t1 cell line, Ultrasonics 54~(4) (2014) 981--989.

\bibitem{wang2015anti}
P.~Wang, C.~Li, X.~Wang, W.~Xiong, X.~Feng, Q.~Liu, A.~W. Leung, C.~Xu,
  Anti-metastatic and pro-apoptotic effects elicited by combination
  photodynamic therapy with sonodynamic therapy on breast cancer both in vitro
  and in vivo, Ultrasonics sonochemistry 23 (2015) 116--127.

\bibitem{kozabaranov2020piezoceramic}
R.~Kozabaranov, V.~Borisenok, I.~Didenkulov, A.~Burkatsky, A.~Egorov,
  D.~Litvinov, V.~Chernov, A piezoceramic resonator for sonoluminescence
  analysis, Acoustical Physics 66 (2020) 262--267.

\bibitem{derosa2002photosensitized}
M.~C. DeRosa, R.~J. Crutchley, Photosensitized singlet oxygen and its
  applications, Coordination Chemistry Reviews 233 (2002) 351--371.

\bibitem{suslick2008inside}
K.~S. Suslick, D.~J. Flannigan, Inside a collapsing bubble: sonoluminescence
  and the conditions during cavitation, Annu. Rev. Phys. Chem. 59 (2008)
  659--683.

\bibitem{ivanov2000one}
A.~V. Ivanov, A.~V. Reshetnickov, G.~V. Ponomarev, One more pdt application of
  chlorin e6, in: Optical Methods for Tumor Treatment and Detection: Mechanisms
  and Techniques in Photodynamic Therapy IX, Vol. 3909, International Society
  for Optics and Photonics, 2000, pp. 131--137.

\bibitem{tronson2002comparison}
R.~Tronson, M.~Ashokkumar, F.~Grieser, Comparison of the effects of
  water-soluble solutes on multibubble sonoluminescence generated in aqueous
  solutions by 20-and 515-khz pulsed ultrasound, The Journal of Physical
  Chemistry B 106~(42) (2002) 11064--11068.

\bibitem{price2004sonoluminescence}
G.~J. Price, M.~Ashokkumar, F.~Grieser, Sonoluminescence quenching of organic
  compounds in aqueous solution: Frequency effects and implications for
  sonochemistry, Journal of the American Chemical Society 126~(9) (2004)
  2755--2762.

\bibitem{tronson2012behavior}
R.~Tronson, M.~F. Tchea, M.~Ashokkumar, F.~Grieser, The behavior of acoustic
  bubbles in aqueous solutions containing soluble polymers, The Journal of
  Physical Chemistry B 116~(46) (2012) 13806--13811.

\bibitem{yasui2002influence}
K.~Yasui, Influence of ultrasonic frequency on multibubble sonoluminescence,
  The Journal of the Acoustical Society of America 112~(4) (2002) 1405--1413.

\bibitem{kazachek2020using}
M.~Kazachek, T.~Gordeychuk, Using photon correlation counter for determining
  the amount of emitting bubbles and number of photons per flash of multibubble
  sonoluminescence, Technical Physics Letters 46 (2020) 263--267.

\bibitem{visheratina2019circular}
A.~K. Visheratina, F.~Purcell-Milton, Y.~K. Gun’ko, A.~Orlova, Circular
  dichroism spectroscopy as a powerful tool for unraveling assembly of chiral
  nonluminescent aggregates of photosensitizer molecules on nanoparticle
  surfaces, The Journal of Physical Chemistry A 123~(37) (2019) 8028--8035.

\bibitem{jain2021fluorescence}
R.~Jain, R.~Pradhan, S.~Hejmady, G.~Singhvi, S.~K. Dubey, Fluorescence-based
  method for sensitive and rapid estimation of chlorin e6 in stealth liposomes
  for photodynamic therapy against cancer, Spectrochimica Acta Part A:
  Molecular and Biomolecular Spectroscopy 244 (2021) 118823.

\end{thebibliography}

\end{document}